\documentclass[11pt,a4paper]{article}
\usepackage[utf8]{inputenc}
\usepackage[T1]{fontenc}
\usepackage{geometry}
\usepackage{graphicx}
\usepackage{booktabs}
\usepackage{url}
\usepackage{hyperref}
\usepackage{stfloats}

\usepackage{comment}
\usepackage{enumitem}
\usepackage{xcolor}
\usepackage{amsmath}

\title{TidyVoice 2026 Challenge Evaluation Plan}

\author{
Aref Farhadipour$^1$, Jan Marquenie$^2$, Srikanth Madikeri$^1$, Teodora Vukovic$^1$, Volker Dellwo$^1$, \\
Kathy Reid$^3$, Francis M. Tyers$^{3,4}$, Ingo Siegert$^2$, Eleanor Chodroff$^1$\\
\\
$^1$Department of Computational Linguistics, University of Zurich, Switzerland\\
$^2$Mobile Dialog Systems, Otto-von-Guericke-University Magdeburg, Germany\\
$^3$Mozilla Foundation, USA\\
$^4$Department of Linguistics, Indiana University, USA\\
\\
\texttt{aref.farhadipour@uzh.ch, jan.marquenie@ovgu.de}
}

\date{\today}

\begin{document}

\maketitle
\footnotetext{Challenge website: \url{https://tidyvoice2026.github.io/}}

\begin{abstract}
    
  The performance of speaker verification systems degrades significantly under language mismatch, a critical challenge exacerbated by the field's reliance on English-centric data. To address this, we propose the TidyVoice Challenge for cross-lingual speaker verification. The challenge leverages the TidyVoiceX dataset from the novel TidyVoice benchmark, a large-scale, multilingual corpus derived from Mozilla Common Voice, and specifically curated to isolate the effect of language switching across approximately 40 languages. Participants will be tasked with building systems robust to this mismatch, with performance primarily evaluated using the Equal Error Rate (EER) on cross-language trials. By providing standardized data, open-source baselines, and a rigorous evaluation protocol, this challenge aims to drive research towards fairer, more inclusive, and language-independent speaker recognition technologies, directly aligning with the Interspeech 2026 theme, "Speaking Together."
  
\end{abstract}

% ----------------------------------------------------------------------

\section{Introduction}

The TidyVoice 2026 Challenge is an evaluation for cross-lingual speaker verification, organized as part of Interspeech 2026. It targets the open problem of speaker verification under language mismatch.

The main objectives are:
\begin{itemize}[leftmargin=1.5em]
    \item measure system performance under controlled cross-language conditions;
    \item provide a simple, reproducible benchmark with clear protocols and metrics;
    \item enable research on language-robust speaker embeddings using a public dataset.
\end{itemize}

Participation is open to all teams able to comply with the rules in this plan.

% ----------------------------------------------------------------------
\section{Task Description}

\subsection{Task Definition}

The task for the TidyVoice Challenge is speaker verification: given a test segment and a target individual's enrollment data, automatically determine whether the target individual is present in the test segment. The test segment along with the enrollment segment(s) from a designated target individual constitute a trial. The system is required to process each trial independently and output a similarity score for that trial. While the score can be in the form of a log-likelihood ratio (LLR), any similarity score that allows threshold-based decision making is acceptable. The LLR for a given trial including a test segment $s$ is defined as

\begin{equation} \label{eq:llr}
\text{LLR}(s) = \log \left( \frac{P(s|H_0)}{P(s|H_1)} \right)
\end{equation}

where $P(\cdot)$ denotes the probability density function (pdf), and $H_0$ and $H_1$ represent the null (i.e., the target individual is present in $s$) and alternative (i.e., the target individual is not present in $s$) hypotheses, respectively.

\subsection{Training Condition}

The training condition is defined as the amount of data/resources used to build a speaker recognition system. The task described above is evaluated over an open training condition.

\textbf{Open Training Condition} -- The open training condition removes limitations on data usage. Participants can use:
\begin{itemize}[leftmargin=1.5em]
    \item The official TidyVoiceX training partition (see Section~\ref{sec:data})
    \item Any other proprietary and/or publicly available data
    \item Publicly available, pre-trained models (e.g., models trained on VoxCeleb~\cite{nagrani2017voxceleb,chung2018voxceleb2}, VoxBlink~\cite{lin2024voxblink2}, self-supervised learning models such as wav2vec2, HuBERT, WavLM, etc.)
    \item External, non-speech data for data augmentation (e.g., noise or reverberation from public corpora like MUSAN~\cite{snyder2015musan})
\end{itemize}

However, there is a strict restriction regarding the source corpus: the only data permitted from the Mozilla Common Voice (MCV) dataset is the official TidyVoiceX training partition. The use of any other data from the MCV corpus is strictly forbidden. The official training data list will be provided on the challenge website.

Participating teams must provide a sufficient description of all data resources (speech and non-speech) as well as pre-trained models used during the training and development of their systems (see Section~\ref{sec:system_description}).

\subsection{Enrollment Conditions}

The enrollment condition is defined as the number of speech segments provided to create a target speaker model. In the TidyVoice Challenge, enrollment data consists of audio segments from a target speaker. The enrollment segments may vary in duration and may contain speech in one or more languages. The specific enrollment format will be provided in the trial files.

\subsection{Test Conditions}

The test conditions in the TidyVoice Challenge are as follows:

\begin{itemize}[leftmargin=1.5em]
    \item The speech duration of the test segments will vary, typically ranging from a few seconds to approximately 60 seconds.
    \item Trials will be conducted with test segments spoken in both the same and different languages as the enrollment segment(s).
    \item \textbf{Important:} The final evaluation set uses different languages from the training and development sets. Specifically, the training and development data contain 40 languages, while the final evaluation set contains 38 unseen languages that are not present in the training and development data. This design tests the system's ability to generalize to completely new languages.
    \item The evaluation phase includes two trial lists:
    \begin{itemize}
        \item \textbf{tv26\_eval-A}: Enrollment utterances from \textbf{seen languages} (languages present in the training and development data) and test utterances from \textbf{unseen languages} (languages not present in the training and development data).
        \item \textbf{tv26\_eval-U}: Both enrollment and test utterances from \textbf{unseen languages} only. This list includes 38 unseen languages that are not present in the training and development data.
    \end{itemize}
    \item Information about the language of each utterance will not be disclosed during evaluation to ensure fair assessment of language-independent systems.
    \item There will be no cross-gender trials.
\end{itemize}

% ----------------------------------------------------------------------
\section{Performance Metrics}

System performance will be ranked based on the following metrics, calculated from the submitted scores:

\subsection{Primary Metric}

The primary metric for ranking will be the Equal Error Rate (EER). Evaluations will be reported on both pooled and four types of trial pair subsets to assess overall robustness and language-specific performance. In the evaluation phase, we have two different trial pair lists that all participants must evaluate and submit results for both trial lists.

The four types of trial pairs in the development phase are:
\begin{itemize}[leftmargin=1.5em]
    \item Target pairs (same speaker, same language)
    \item Target pairs (same speaker, different languages)
    \item Non-target pairs (different speakers, same language)
    \item Non-target pairs (different speakers, different languages)
\end{itemize}

\subsection{Secondary Metric}

The secondary metric is the Minimum Detection Cost Function (minDCF), which provides a cost-weighted combination of false acceptance and false rejection errors. It is defined as:

\begin{equation}
\text{DCF} = C_{\text{miss}} \times P_{\text{miss}} \times P_{\text{tar}} + C_{\text{fa}} \times P_{\text{fa}} \times (1 - P_{\text{tar}})
\end{equation}

where $C_{\text{miss}}$ and $C_{\text{fa}}$ are the costs of a miss and a false alarm, respectively; $P_{\text{miss}}$ and $P_{\text{fa}}$ are the corresponding error probabilities; and $P_{\text{tar}}$ is the prior probability of a target trial. The minimum value of this function over all possible decision thresholds is reported as \textit{minDCF}.

\textbf{Evaluation Setting:} We set $C_{\text{miss}} = C_{\text{fa}} = 1.0$ and $P_{\text{tar}} = 0.01$. This configuration serves as our evaluation setting and is used as a secondary metric for tie-breaking and comprehensive performance analysis.

% ----------------------------------------------------------------------
\section{Rules and Requirements}

To ensure a fair and standardized evaluation, all participants must adhere to the following rules and requirements:

\begin{itemize}[leftmargin=1.5em]
    \item \textbf{Training Data Regulations:} This is an open-condition challenge, and participants are permitted to use any public or private datasets to train their systems. The use of all non-challenge data must be fully disclosed in the system description paper. However, there is a strict restriction regarding the source corpus: the only data permitted from the MCV dataset is the official TidyVoiceX training partition, as defined in Section~\ref{sec:data}. The official training data list will be provided on the challenge website.

    \item \textbf{Prohibition of Re-identification and Data Recombination:} Any attempt to determine the real-world identity of a speaker, or to link speakers in the challenge data to individuals or records in \emph{external} datasets, services, or metadata (``data recombination'') is strictly forbidden. This includes (but is not limited to) voiceprint matching against external corpora, cross-dataset record linkage, or the use of personally identifying metadata. Violations result in disqualification and notification to the organizers and hosting institutions.

    \item \textbf{Scope of Tasks: Verification Only:} This challenge evaluates \emph{speaker verification} (same/different identity) on the provided splits. \emph{Speaker identification} (closed-set or open-set classification among $K$ speakers) is \emph{out of scope} and will not be scored.

    \item \textbf{Data Integrity:} Manual correction or re-labeling of the officially provided challenge data is strictly prohibited.

    \item \textbf{Use of Pre-trained Models:} The use of publicly available, pre-trained models (e.g., models trained on VoxCeleb~\cite{nagrani2017voxceleb,chung2018voxceleb2}, VoxBlink~\cite{lin2024voxblink2}, self-supervised learning models such as wav2vec2, HuBERT, WavLM, etc.) is permitted and encouraged, provided their use is explicitly and thoroughly declared in the system description paper.

    \item \textbf{Data Augmentation:} The use of external, non-speech data for data augmentation (e.g., noise or reverberation from public corpora like MUSAN~\cite{snyder2015musan}) is permitted and encouraged, but has to be explicitly and thoroughly disclosed in the system description paper.

    \item \textbf{Submission Limit:} Each participating team is allowed to submit scores for a maximum of three distinct systems per trial list.

    \item \textbf{Eligibility for Final Ranking:} To be eligible for inclusion in the final ranking and challenge results, participants must submit a system description paper to the dedicated Interspeech 2026 special session.

    \item \textbf{Reproducibility \& Artifacts:} The top-performing team in each task must submit their trained model and an inference script \emph{in a single \texttt{.zip} archive}, sufficient to reproduce the reported scores and prevent post-submission alterations (e.g., manual correction); if results cannot be reproduced, the team will be disqualified and this requirement will pass to the next highest-ranked team.
\end{itemize}

% ----------------------------------------------------------------------
\section{Evaluation Protocol}

To facilitate information exchange between the participants and organizers, all evaluation activities are conducted over web-based platforms.

\subsection{Challenge Registration}

Participants must register for the challenge through the official registration form available on the challenge website. Registration is required to access the evaluation datasets and submit results.

\subsection{Challenge Phases}

The TidyVoice Challenge consists of two main phases:

\textbf{Development Phase:} This phase is conducted offline by participants. During this phase, participants use the provided training and development datasets to develop, train, and tune their systems locally. Participants can experiment with different approaches and evaluate their systems on the development set without any online submission.

\textbf{Evaluation Phase:} This phase involves online ranking on the CodaBench website. During the evaluation phase, participants will submit their results on the development set through the CodaBench platform. The link to the challenge on CodaBench will be shared with participants during the evaluation phase. Rankings will be determined based on performance on the development set, and participants can view their position on the public leaderboard.

\subsection{Submission Requirements}

Each team must make at least one valid submission for the evaluation set, processing all test segments. Submissions with missing test segments will not pass the validation step, and hence will be rejected.

The participants can register up to three systems per trial list. This results in a maximum of three registered systems per trial list, one submission per system. Bug-fixes do not count toward this limit.

Each team is required to submit system descriptions at the designated time (see Section~\ref{sec:schedule}). The evaluation results are made available only after the system description report is received and confirmed to comply with guidelines described in Section~\ref{sec:system_description}.

\subsection{System Output Format}
\label{sec:output_format}

The system output file is composed of a set of records where each record contains a trial given in the trial file and one additional column specifying a similarity score output by the system for the trial. The order of the trials in the system output file must follow the same order as the trial list. Each record is a single line containing three fields separated by tabs. There should be one output file for each trial list for each system. A validation script will be provided to participants to ensure their submission files conform to the required format before the final deadline.

\textbf{Trial File Format:} The evaluation trial list will be provided as a tab-separated text file. Each line in this file represents a single trial pair to be scored. The format for each line is:

\vspace{0.5em}
\noindent\parbox{\columnwidth}{
    \texttt{enrollment\_file\ \ \ test\_file}
}

\textit{Example:}

\noindent\parbox{\columnwidth}{
    \texttt{spk\_A\_enroll.wav\ \ \ test\_001.wav}\\
    \texttt{spk\_A\_enroll.wav\ \ \ test\_002.wav}
}

\textbf{System Output Format:} Participants must submit a single output file (.txt format) containing a score for every trial specified in the trial file. The output file must follow the same order as the trial file and add a third column with the similarity score. The format for each line is:

\vspace{0.5em}
\noindent\parbox{\columnwidth}{
    \texttt{enrollment\_file\ \ \ test\_file\ \ \ score}
}

\textit{Example:}

\noindent\parbox{\columnwidth}{
    \texttt{spk\_A\_enroll.wav\ \ \ test\_001.wav\ \ \ 0.862}\\
    \texttt{spk\_A\_enroll.wav\ \ \ test\_002.wav\ \ \ 0.124}
}

\subsection{System Description Format}
\label{sec:system_description}

Each team is required to submit a system description for each system (all primary and alternative systems). A system description must include the following items:

\begin{itemize}[leftmargin=1.5em]
    \item A complete description of the system components, including front-end (e.g., speech activity detection, features, normalization) and back-end (e.g., background models, speaker embedding extractor, LDA/PLDA) modules along with their configurations (i.e., filterbank configuration, dimensionality and type of the acoustic feature parameters, as well as the acoustic model and the backend model configurations).
    
    \item A complete description of the data partitions used to train the various models (as mentioned above). Teams are encouraged to report how having access to the Development set impacted the performance.
    
    \item A complete description of all external data resources (speech and non-speech) as well as pre-trained models used during the training and development of their systems.
    
    \item Performance of the submissions for that system on the TidyVoice Challenge Development set (or a derivative/custom dev set), using the scoring software provided via the web platform. Teams are encouraged to quantify the contribution of their major system components that they believe resulted in significant performance gains.
    
    \item A report of the CPU (single threaded) and GPU execution times as well as the amount of memory used to process a single trial (i.e., the time and memory used for creating a speaker model from enrollment data as well as processing a test segment to compute the score).
\end{itemize}

The system description should follow the latest IEEE ICASSP conference proceeding template or Interspeech format.

% ----------------------------------------------------------------------
\section{Data}
\label{sec:data}

The TidyVoice Challenge is built upon a curated data partition derived from the Mozilla Common Voice (MCV) corpus~\cite{zhang2025quantifying, hintz2024commonbench}: TidyVoiceX~\cite{farhadi2026tidy}. 

\subsection{Official Challenge Dataset}

The primary and official dataset for this challenge is the TidyVoiceX partition~\cite{farhadi2026tidy}. It serves as the core resource for both training and evaluation. As shown in Table~\ref{tab:dataset_comparison}, TidyVoiceX provides a large-scale, multilingual benchmark specifically designed for cross-lingual speaker verification, featuring a clearly defined training and development split across 40 languages.

\textbf{Important Note:} The final evaluation set uses a different set of languages than the training and development data. While the training and development sets contain 40 languages, the final evaluation set contains 38 \textbf{unseen languages} that do not overlap with the training and development languages. This design ensures that systems are evaluated on their ability to generalize to completely new languages, testing true language-independent speaker verification capabilities.

\begin{table}[h]
\centering
\caption{The official train and development datasets in the TidyVoice Challenge}
\label{tab:dataset_comparison}
\resizebox{0.75\textwidth}{!}{%
\begin{tabular}{l|c|c|c|c|c}
\hline
\textbf{Dataset} & \textbf{\# Spkr} & \textbf{\# Lang} & \textbf{\# Utt} & \textbf{Dur (H)} & \textbf{Domain} \\
\hline
\hline
\textbf{TidyVoiceX} & \textbf{4,474} & \textbf{40} & \textbf{321,711} & \textbf{457} & \textbf{Read}  \\
\quad TidyVoiceX: Train & 3,666 & 40 & 262K & 370 & Read  \\
\quad TidyVoiceX: Dev & 808 & 40 & 60K & 87 & Read  \\
\hline
\textbf{TidyVoiceX2: Eval} & 2,200 & 38 & 200K & 300 & Read  \\
\hline
\end{tabular}%
}
\end{table}

\subsection{Data Characteristics}

The TidyVoiceX dataset adheres to the following principles to ensure fairness, privacy, and reproducibility:

\begin{itemize}[leftmargin=1.5em]
  \item All data features pseudonymized speaker IDs to protect contributor privacy.
  \item The dataset is organized with speakerID folders directly inside each dataset folder, which then contain languageID subfolders with the corresponding audio files for that speaker in that specific language.
  \item Audio files are in standard WAV format with 16 kHz sampling frequency.
  \item The domain is read speech, which minimizes stylistic and phonetic variability, allowing for a more precise analysis of cross-lingual acoustic modeling.
  \item The official test data, along with two trial pair lists for the final evaluation, will be clearly defined and documented on the challenge website.
  \item Information about the language of each utterance will not be disclosed during evaluation.
\end{itemize}

\subsection{Data Restrictions}

\textbf{The use of any other data from the MCV corpus is strictly forbidden.} Only the official TidyVoiceX training partition may be used from the Mozilla Common Voice dataset. The official training data list will be provided on the challenge website.

\subsection{Data Access}

The TidyVoiceX dataset is available through the Mozilla Data Collective. Participants must register for the challenge and obtain access credentials. The dataset package contains both training and development data:

\begin{itemize}[leftmargin=1.5em]
    \item \textbf{TidyVoiceX\_Train/}: Training dataset with multi-lingual speaker recordings
    \item \textbf{TidyVoiceX\_Dev/}: Development dataset for system tuning and validation
    \item Speaker identity labels and language annotations
    \item Audio files in standard WAV format with 16 kHz sampling frequency
    \item Cross-lingual speaker samples across both splits
\end{itemize}

The dataset size is approximately 50 GB. Detailed download instructions, including how to create a Mozilla Data Collective API key and use the provided download script, are available on the challenge website.
% 

% \resizebox{\textwidth}{!}{%
% \begin{tabular}{l l c c c c c c}
% \toprule
% \textbf{Model} & \textbf{Training Data} & \textbf{Vox1-O} & \textbf{Vox1-E} & \textbf{Vox1-H} & \textbf{CANDOR} & \textbf{Tidy-M} & \textbf{Tidy-X} \\
% \midrule
% ResNet-34 & Tidy-M (from scratch) & 11.14 / 0.89 & 11.96 / 0.89 & 15.70 / 0.91 & 4.86 / 0.32 & 0.39 / 0.06 & 2.83 / 0.71 \\
% ResNet-34 & VB2 [5] + Tidy-M & 1.15 / 0.12 & 1.29 / 0.14 & 2.39 / 0.22 & 2.00 / 0.10 & 0.67 / 0.09 & 1.90 / 0.69 \\
% ResNet-293 & VB2 [5] + Tidy-M & 0.53 / 0.06 & 0.83 / 0.08 & 1.54 / 0.14 & 1.60 / 0.07 & 0.35 / 0.19 & 1.65 / 0.71 \\
% \bottomrule
% \end{tabular}%
% }
% \end{table*}

% ----------------------------------------------------------------------
\section{Baseline Systems}

To provide a strong starting point for participants and ensure a fair and reproducible evaluation, we will publicly release an official baseline system. This system represents an approach that uses a pre-trained model fine-tuned on the provided data. The goal is to lower the barrier to entry and allow participants to benchmark their own novel methods against this robust baseline.

The official baseline system is implemented using the WeSpeaker toolkit, a popular and widely-used toolbox for Automatic Speaker Verification (ASV) tasks. By leveraging this established framework, participants can more easily develop new achievements and build upon existing implementations, as the baseline provides a solid foundation within a familiar ecosystem.

\subsection{Baseline Architecture}

The baseline system uses a SimAM-ResNet34 architecture that is:
\begin{enumerate}[leftmargin=1.5em]
    \item \textbf{Pretrained} on VoxBlink2 and VoxCeleb2 datasets
    \item \textbf{Fine-tuned} on the TidyVoiceX training set using large-margin training
\end{enumerate}

\subsection{Baseline Results}

The baseline achieves the following overall performance on the TidyVoiceX development set:

\begin{table}[h]
\centering
\caption{Performance of the official challenge baseline system on the TidyVoiceX development set}
\label{tab:baseline_performance}
\resizebox{0.85\textwidth}{!}{%
\begin{tabular}{l|c|c|c|c|c}
\hline
\textbf{Architecture} & \textbf{Pretraining Data} & \textbf{Fine-tuning Data} & \textbf{EER (\%)} & \textbf{MinDCF} & \textbf{Model} \\
\hline
SimAM-ResNet34 & VoxBlink2 + VoxCeleb2 & TidyVoiceX Train & 3.07 & 0.82 & Available\footnote{\url{https://huggingface.co/areffarhadi/Resnet34-tidyvoiceX-ASV}} \\
\hline
\end{tabular}%
}
\end{table}

\noindent
To better understand the baseline system's behavior, we analyze performance across four distinct trial pair scenarios that capture different combinations of speaker identity and language matching.

\begin{table}[h]
\centering
\caption{Baseline EER (\%) summary by language match (TidyVoiceX dev)}
\label{tab:baseline_trial_categories}
\resizebox{0.80\textwidth}{!}{%
\begin{tabular}{l|l|c}
\hline
\textbf{Target (same speaker)} & \textbf{Non-target (diff speaker)} & \textbf{EER (\%)} \\
\hline
Different language & Different language & 1.79 \\
Different language & Same language & 5.19 \\
Same language & Different language & 0.88 \\
Same language & Same language & 2.97 \\
\hline
\multicolumn{2}{l|}{\textbf{Overall (all target vs all non-target)}} & \textbf{3.07} \\
\hline
\end{tabular}%
}
\end{table}

\noindent
On the final evaluation test set (\textbf{TidyVoiceX2}), we report baseline performance on two trial lists:

\begin{table}[h]
\centering
\caption{Baseline performance on the TidyVoiceX2 evaluation trial lists}
\label{tab:baseline_eval_performance}
\resizebox{0.85\textwidth}{!}{%
\begin{tabular}{l|c|c|c}
\hline
\textbf{Trial List} & \textbf{Description} & \textbf{\# Trials} & \textbf{EER (\%) / minDCF} \\
\hline
\textbf{tv26\_eval-A} & All evaluation languages  & 4.0M & 9.058 / 0.65 \\
\textbf{tv26\_eval-U} & Unseen languages only & 1.280M & 11.59 / 0.60 \\
\hline
\end{tabular}%
}
\end{table}

\noindent
\textbf{Language Dependency Analysis.} The detailed results reveal an important finding: the baseline system exhibits a significant dependency on language information, which should not be a discriminative factor in speaker verification. This is evident from the performance differences across the four scenarios:
\begin{itemize}[leftmargin=1.5em]
    \item \textbf{Best performance (EER: 0.88\%):} When target pairs share the same language and non-target pairs use different languages, the system can leverage language mismatch as a discriminative cue.
    \item \textbf{Worst performance (EER: 5.19\%):} When target pairs use different languages but non-target pairs share the same language, the system struggles because it cannot rely on language differences.
\end{itemize}

This pattern indicates that the system is relying on language characteristics rather than focusing solely on speaker characteristics, which is problematic for cross-lingual speaker verification.

\begin{figure}[htbp]
\centering
\includegraphics[width=0.95\textwidth]{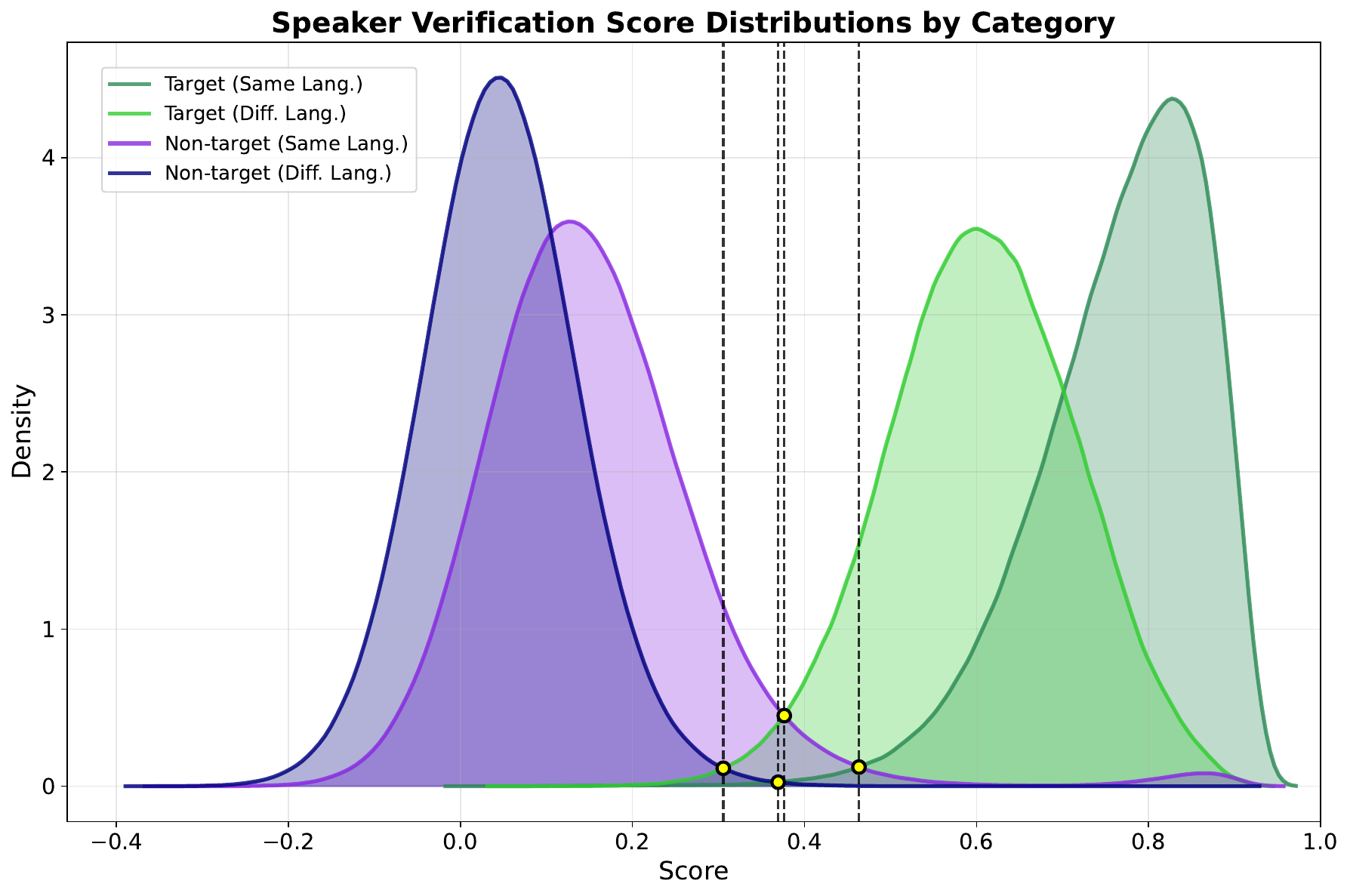}
\caption{Score distributions for the four trial pair categories on the TidyVoiceX development set.}
\label{fig:score_distributions}
\end{figure}

\subsection{Notes}

The baseline results are reported to provide a reproducible reference point and a sanity check for participants. Detailed condition-wise analyses and score distributions are intentionally omitted from this evaluation plan to keep it concise.

\subsection{Resources}

We release the complete training recipe, evaluation scripts, and the pre-trained checkpoint for this model to the public:

\begin{itemize}[leftmargin=1.5em]
    \item \textbf{Baseline Code Repository}\footnote{\url{https://github.com/areffarhadi/wespeaker/tree/master/examples/tidyvocie}}
    \item \textbf{Model Repository}\footnote{\url{https://huggingface.co/areffarhadi/Resnet34-tidyvoiceX-ASV}}
\end{itemize}

% ----------------------------------------------------------------------
\section{Schedule}
\label{sec:schedule}

The following is the schedule for the TidyVoice Challenge:

\begin{center}
\begin{tabular}{ll}
\toprule
\textbf{Milestone} & \textbf{Date}\\
\midrule
Development Phase & December 1, 2025 \\
Evaluation Phase & January 25, 2026 \\
Registration Deadline & February 5, 2026 \\
Result Submission Deadline & February 10, 2026 \\
Release Selected Teams & February 12, 2026 \\
System Description Deadline & February 25, 2026 \\
\bottomrule
\end{tabular}
\end{center}

We will encourage participants to start early on the paper writing; many sections can be written before obtaining the final evaluation results (e.g., background, system description, etc.).

\bibliographystyle{IEEEtran}
\bibliography{mybib}

\end{document}